\documentclass[aps,prm,twocolumn,floatfix,superscriptaddress]{revtex4}
\usepackage{times}
\usepackage{graphicx}
\usepackage{epsfig}
\usepackage{amsmath}
\usepackage{amssymb}
\usepackage[version=3]{mhchem}
\usepackage{multirow}

\begin{document}

\title{On the polytypism of layered {\it MX}$_{\boldsymbol2}$ materials}

\author{Emma H. Wolpert}
\affiliation{Department of Chemistry, University of Oxford, Inorganic Chemistry Laboratory, South Parks Road, Oxford OX1 3QR, U.K.}
\affiliation{Department of Chemistry, Imperial College London, SW7 2AZ, U.K.}

\author{Simon J. Cassidy}
\affiliation{Department of Chemistry, University of Oxford, Inorganic Chemistry Laboratory, South Parks Road, Oxford OX1 3QR, U.K.}

\author{Andrew L. Goodwin}
\email{andrew.goodwin@chem.ox.ac.uk}
\affiliation{Department of Chemistry, University of Oxford, Inorganic Chemistry Laboratory, South Parks Road, Oxford OX1 3QR, U.K.}

\date{\today}
\begin{abstract}
We revisit the problem of polytypism in layered {\it MX}$_2$ materials, with a view to reinterpreting the phase space accessible to this family. Our starting point is to develop a simple, constructive and compact label for the most commonly observed stacking arrangements that is similar to the Glazer notation used to label tilt systems in perovskites. The key advantage of this label in the context of {\it MX}$_2$ systems is that it contains sufficient information to generate the corresponding stacking sequences uniquely. Using a related approach, we generate a Cartesian representation of the phase space containing all possible {\it MX}$_2$ polytypes, with the most common structures appearing as limiting cases. We argue that variation in \emph{e.g.}\ composition, or temperature, or pressure may allow navigation of this phase space along continuous paths. This interpretation is shown to be consistent with the structural evolution of stacking-faulted {\it MX}$_2$ systems as a function of temperature and composition. Our study highlights the potential for controlling composition/structure/property relationships amongst layered {\it MX}$_2$ materials in ways that might not previously have been obvious.
 \end{abstract}


\maketitle

\section{Introduction}

Layered {\it MX}$_2$ materials are a long-studied and broad family of particular currency in the fields of topological insulators, thermoelectrics, and strongly-correlated electronic materials. Topical examples include MoS$_2$ \cite{Radisavljevic_2011,LopezSanchez_2013,VazquezSulleiro_2022}, PtBi$_2$ \cite{Gao_2017, Feng_2019,Xing_2020}, and WTe$_2$ \cite{Ali_2014,Fei_2017,Pan_2022}. Central to an understanding of their electronic properties is an appreciation of the underlying atomic-scale structure, which can be particularly complex for this family.

The structures of individual {\it MX}$_2$ layers are usually simple enough: they are, in general, comprised of edge-sharing octahedra or trigonal prisms \cite{Kertesz_1984}. Complexity arises because of the very large number of different possible stacking sequences that in turn give very different crystal symmetries (\emph{e.g.}\ polar, chiral, centrosymmetric) \cite{Palosz_1983,Palosz_1983b}. Because inter-layer interactions are inherently weak, the energy landscape associated with stacking variations can be shallow indeed, and it is often possible to isolate polymorphs with different stacking sequences for a single common composition. Such stacking-sequence polymorphs are called polytypes, and (by way of example) CdI$_2$ is reported to exhibit more than 150 of them \cite{Palosz_1983}.

There is a long and interesting history of the study of polytypism in {\it MX}$_2$ systems. On the one hand, there has been a sustained experimental effort to identify as many polytypes for different {\it MX}$_2$ chemistries as possible, and then to categorise and/or label these polytypes in various empirical ways \cite{Rao_1981,Salje_1987}. On the other hand, statistical mechanical models have been used to address the underlying physical origin of complexity in the family \cite{Hazen_1981,Price_1983,Yeomans_1986}. It was realised early on that problems of layer stacking often map onto one-dimensional (1D) Ising models \cite{Elliott_1961,Smith_1984,Price_1984}. A famous result in the field concerns the so-called ANNNI model, which takes into account different---and perhaps competing---interactions between nearest-neighbour and next-nearest-neighbour layers \cite{Elliott_1961}. The ANNNI phase diagram admits an infinity of different phases corresponding to stacking sequences with arbitrarily long repeat lengths, and at face value might have explained the complexity observed experimentally. Unfortunately, there is no general correspondence between experimentally-observed polytypes and the ANNNI phases.

Here we revisit the problem of polytypism in layered {\it MX}$_2$ materials by focusing on correlations within stacking sequences, rather than the interactions from which they might arise. We have a few particular objectives. The first is to develop a simple, constructive and compact label for the most commonly observed stacking arrangements---\emph{i.e.}\ a label which contains sufficient information to generate the corresponding stacking sequences uniquely. Modelling our approach on the Glazer notation used to describe tilt systems in perovskites, we hope that enumerating the stacking sequences using constructive labels will allow for targeted explanation of different stacking sequences \cite{Woodward_1997}. The second goal is to generate a representation of the phase space that contains all possible {\it MX}$_2$ polytypes, with the most common structures appearing at its boundary. Our hope here is that variation in \emph{e.g.}\ composition, or temperature, or pressure might allow navigation of this phase space along continuous paths. And, third, we reinterpret the diffraction patterns of various ostensibly complex polytypes in terms of disordered stacking arrangements found within the interior of this {\it MX}$_2$ phase space. In addressing these points, our study highlights the potential for controlling composition/structure/property relationships amongst layered {\it MX}$_2$ materials in ways that might not previously have been obvious.

\begin{figure*}
\begin{center}
\includegraphics{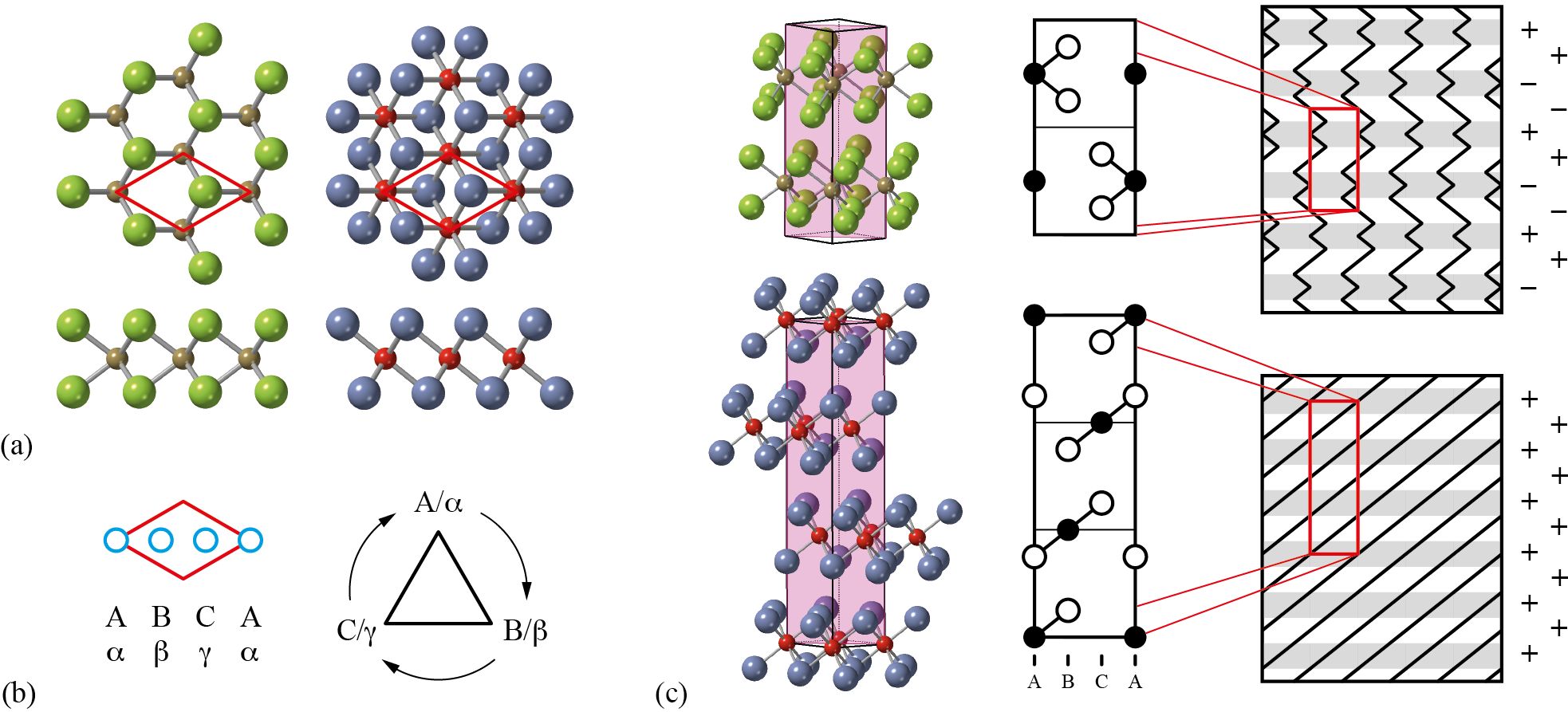}
\end{center}
\caption{\label{fig1} Fundamental structural characteristics and representations of layered {\it MX}$_2$ polytypes. (a) Individual layers comprise close-packed {\it M} centres in either trigonal prismatic (left) or octahedral (right) coordination. {\it X} anions are shown as large spheres; {\it M} cations as small spheres. The unit cells of these arrangements are shown in red outline. Views are given along (top) and perpendicular to (bottom) the layer normal. (b) {\it M} and {\it X} positions can be assigned to one of three sites within the underlying (projected) unit cell, denoted using the Greek letters $\alpha,\beta,\gamma$ for the former, and the Roman letters A,B,C for the latter. Individual layers can be assigned a handedness according to the sense of permutation of these positional labels when traversing the layer from the bottom up (octahedral) or from the inside out (trigonal prismatic). (c) Representations of the MoS$_2$ (top) and CdCl$_2$ (bottom) structures with crystallographic unit-cells shown in black outline. In each case, the central $y=1-x$ plane is shown in pink. The positions of atoms lying on this plane provides a two-dimensional representation of the structure types, shown in the central panel using black filled circles to denote {\it M} atoms and open circles to denote {\it X} atoms. The right-hand panels show a concise visual representation of polytype structure (used \emph{e.g.}\ in Ref.~\citenum{Palosz_1983b}) which gives the orientation within (gray areas) and between (white areas) layers. The corresponding handednesses are shown to the right of each panel as `$+$' or `$-$' symbols. Note that in the MoS$_2$ structure, both layer orientation and slip orientation alternate from layer to layer. By contrast, in CdCl$_2$ all layers and all slip directions have a single common handedness.}
\end{figure*}

Our paper is arranged as follows. We begin with a short review of the fundamental crystallography of layered {\it MX}$_2$ polytypes. In doing so, we revisit the various key nomenclatures used historically to distinguish different structure types. We then introduce our new approach to labelling (and understanding) polytypes in terms of correlations between successive layer and inter-layer geometries. This approach suggests a spatial organisation of {\it MX}$_2$ polytypes into three-dimensional phase fields, and we proceed first to introduce and subsequently to explore this mapping. By considering key points and trajectories within these phase fields, we determine the effect of different correlations on X-ray diffraction patterns. We then relate our calculations to previously-published experimental data, demonstrating that the complexity often attributed to long-period stacking sequences might be better understood in terms of disordered stacking arrangements. Our paper concludes with a discussion of the implications of our results for polytype characterisation and polytype selection in {\it MX}$_2$ systems more generally.

\section{Background: geometric considerations}

The individual layers of {\it MX}$_2$ systems consist of a close-packed array of {\it M} cations at their core, with close-packed arrays of {\it X} anions above and below [Fig.~\ref{fig1}(a)]. The {\it X} anions always sit above or below holes in the {\it M} layer. If the coordination geometry of {\it M} is trigonal prismatic, then the two {\it X} layers sit directly above one another; for octahedral geometries, the two {\it X} layers are slipped relative to each other. The symmetries of the two arrangements are captured formally by the layer groups $p\bar6m2$ and $p\bar3m1$, respectively. Electronically-driven distortions (\emph{e.g.}\ {\it M}--{\it M} bond formation, Peierls instabilities, charge-density wave formation) can break the symmetry of individual layers; but we need not consider these additional symmetry-lowering effects here, since they are orthogonal to the polytypism problem.

Both octahedral and trigonal prismatic layer symmetries are lower than that of the central close-packed array of {\it M} atoms ($p6/mmm$), and this gives rise to two distinguishable but equivalent layer orientations in each case. It will be convenient to assign to each orientation a `handedness'. To do so, the labels A/$\alpha$, B/$\beta$ and C/$\gamma$ are assigned, respectively, to the coordinates $(0,0)$, $(\frac{1}{3},\frac{2}{3})$ and $(\frac{2}{3},\frac{1}{3})$ within the two-dimensional projection of the $p6/mmm$ layer unit-cell; the Roman characters are used for {\it X} anion positions, and the Greek characters for {\it M} cation positions. Then, progressing from the bottom {\it X} layer to the top {\it X} layer (octahedral coordination), or from the central {\it M} layer outwards (trigonal prismatic coordination), the ion positions correspond to permutations of either the clockwise cyclic sequence A/$\alpha\rightarrow$B/$\beta\rightarrow$C/$\gamma(\rightarrow$A/$\alpha)$ or its anticlockwise inverse A/$\alpha\rightarrow$C/$\gamma\rightarrow$B/$\beta(\rightarrow$A/$\alpha)$ [Fig.~\ref{fig1}(b)]. The sequences A$\beta$C and B$\gamma$B are examples denoting clockwise octahedral and anticlockwise trigonal prismatic coordination geometries, respectively.

In forming different polytypes, individual {\it MX}$_2$ layers stack one above the other. Irrespective of the coordination geometry of {\it M} cations, these stacking arrangements are governed by a single, universally-observed rule: namely, that the anion layers of neighbouring stacks occupy different positions relative to the underlying close-packed unit-cell. In other words, an A$\beta$C layer might be followed by a B$\alpha$C layer, but not by a C$\beta$A layer---the sequence A$\beta$C C$\beta$A being too unfavourable. Since the anion layers bordering each van der Waals gap are always in different positions, the slip direction from layer to layer can also be assigned a handedness as if it were an octahedrally-coordinated {\it MX}$_2$ layer of its own. Taking the sequence A$\beta$C A$\beta$C as an example, the slip direction from C to A is assigned the anticlockwise handedness of the associated octahedral layer sequence C$(\beta)$A.

It is not surprising, therefore, that layered {\it MX}$_2$ materials can be so structurally complex: there are two possible {\it M} coordination geometries for each layer, two possible layer orientations for each geometry, and two possible slip directions as each new layer is added to the last. In principle, an {\it MX}$_2$ solid containing $N$ layers has $\sim8^N$ possible polytypes. In practice, a handful of high-symmetry arrangements tend to occur most frequently; some key examples are illustrated in Fig.~\ref{fig1}(c). Here we include also frequently-used representations of AB$_2$ structures that derive from taking a cut through the $y=1-x$ plane of the three-dimensional (3D) unit cell. The positions of {\it M} and {\it X} atoms on this cut allow the 3D structure to be uniquely described in two dimensions, facilitating comparison from polytype to polytype.

\section{Stacking sequence notation}

There are remarkably many different labelling systems used in different communities to describe polytype structures. The so-called `ABC' notation we have already introduced provides an explicit description of layer positions that is exact but cumbersome \cite{Palosz_1982}. H{\"a}gg notation is a more compact variant that condenses layer and inter-layer arrangements into `$+$' and `$-$' symbols, denoting clockwise and anticlockwise sequences, respectively \cite{Hagg_1943}. Zhdanov compresses H{\"a}gg's notation further \cite{Zhdanov_1945}, giving only the number of consecutive layers of common handedness. The `hc', `xyz', and `t--o--f' notations are further alternatives that can also be used to transcribe layer arrangements in a condensed form \cite{Palosz_1980}. These various different notations are all flexible in the sense that they can be used to denote \emph{any} polytype of a given, fixed, {\it M} coordination geometry. Likewise they are constructive in the sense that there is a unique mapping (up to origin shift) from label to stacking sequence.

Yet none of these notations is particularly common in contemporary literature. Instead there is a preference for polytype labels---\emph{e.g.}\ `$1T$', `$2H_a$', `$6R$', $\ldots$---which simply combine the number of layers within the unit cell together with the crystal symmetry (Ramsdell symbols \cite{Ramsdell_1945}). Here there are a number of problems. First, different communities use different labels for the same structure, or the same label for different structures: a frequently-encountered example is that of the textbook CdI$_2$ structure type, which is sometimes denoted `$1T$' and other times `$2H$' \cite{Minagawa_1978}. Second, different stacking sequences can result in unit cells with the same size and crystal system---hence the awkwardness of the $2H_a$, $2H_b$, and $2H_c$ labels, and the ambiguity of labels such as $12R$. And, third, there is generally no means of determining a stacking sequence from a label; instead one simply has to learn which polytype is implied in each case. The key advantage---and probably the reason for its widespread use---is that commonly-occurring polytypes are labelled succinctly.

Recognising the tension between precision, on the one hand, and ease, on the other hand, we were reminded of the success of Glazer notation used to denote different octahedral tilt combinations in perovskites \cite{Glazer_1972}. Glazer symbols such as $a^+b^-b^-$ are precise in that they uniquely describe the particular combination of tilts governing the symmetry of a perovskite (in this case, in-phase tilts along the $\mathbf a$ axis and equal-magnitude out-of-phase tilts along $\mathbf b$ and $\mathbf c$). They are helpful too in making clear the relationships between different phases \cite{Howard_1992}, and hence rationalising transitions between different tilt systems. In proposing his nomenclature, Glazer knew that it could not label all theoretically-possible tilt systems \cite{Howard_2008}, but its success is that it focuses on the tilt systems most frequently observed in practice: those propagating either in-phase (zone-centre) or out-of-phase (zone-boundary).

We suggest that a similar approach might be taken with regards to polytypism in AB$_2$ structures. Amongst the $8^N$ possible layer sequences, it is those for which layer orientations and slip directions propagate in-phase or out-of-phase that are most commonly observed.

Three components must be considered: the relationship between successive layer orientations, the relationship between successive slip orientations, and the coupling between the two. Hence, following Glazer's lead, we propose a three-component symbol of the form $l^\pm s^\pm c^\pm$, denoting in turn the layer ($l$), slip ($s$) and coupling ($c$) correlations. We further use the letters $o,t$ in the $l$ position to denote octahedral and trigonal-primsatic coordination geometries, respectively, which we take to be uniform amongst a single phase. So, the symbol $o^+s^+c^+$ describes a polytype in which octahedrally-coordinated layers have the same (\emph{i.e.}\ in-phase) orientation and slip direction from layer to layer; this is the CdCl$_2$ structure type [Fig.~\ref{fig1}(c)]. Likewise, the MoS$_2$ structure type is given by $t^-s^-c^-$: it consists of trigonal-prismatic layers whose handedness inverts from layer to layer (\emph{i.e.}\ out-of-phase)---as does the slip direction. The coupling term $c^-$ implies that the handedness of a layer orientation is opposite to that of the slip direction directly above it [Fig.~\ref{fig1}(c)].

It is straightforward then to enumerate all polytypes with Glazer-like labels. There can be at most 16 $(=2^4)$, since there are two coordination geometries, and two phase choices for each of layer orientations, slip directions, and coupling terms. In practice---as with Glazer tilts---not all combinations are physically realisable. An example of a `forbidden' label is $o^+s^-c^+$: since the layer orientations are always the same, but the slip direction alternates from van der Waals gap to van der Waals gap, the coupling between layer orientation and slip direction cannot always be in-phase, as implied. Rather, the combination of $o^+$ and $s^-$ implies $c^0$, and so the correct label is $o^+s^-c^0$. Accounting for other such forbidden cases, there remain in total 12 unique allowed labels, which we summarise in Table~\ref{table1}. Of these 12, there are two pairs of two that correspond to opposite polarities of the same structure type, giving 10 symmetry-distinct polytypes that include all of the most common candidates. Full structural descriptions of all polytypes are provided in the Appendix.

\begin{table}
\caption{Glazer labels for key {\it MX}$_2$ polytypes.\label{table1}}
\begin{tabular}{l|cccccc}\hline

\multirow{2}{*}{Symbol}&\multirow{2}{*}{$Z$}&\multirow{2}{*}{Space group}&Ramsdell&&\multirow{2}{*}{Example}&\multirow{2}{*}{Ref.}\\
& & &symbol(s)&&&\\ \hline
$o^+s^+c^+$&3&$R\bar3m$&$3R$, $6R$&&CdCl$_2$&\cite{Pauling_1929}\\
$o^+s^+c^-$&1&$P\bar3m1$&$1T$, $2H$&&CdI$_2$&\cite{Bozorth_1922}\\
$o^+s^-c^0$&6&$R\bar3m$&$6R$&\\
$o^-s^+c^0$&6&$R\bar3m$&$6R^\prime, 12R$&&PbI$_2$ (high $T$)&\cite{Minagawa_1981}\\
$o^-s^-c^+$&2&$P6_3mc$&$1T^\prime, 4H (\uparrow)$&\multirow{2}{*}{$\bigg\}$}&\multirow{2}{*}{CdI$_2$ (high $T$)}&\multirow{2}{*}{\cite{Minagawa_1976}}\\
$o^-s^-c^-$&2&$P6_3mc$&$1T^\prime, 4H (\downarrow)$&\\ \hline

$t^+s^+c^+$&3&$R3m$&$3R (\uparrow)$&\multirow{2}{*}{$\bigg\}$}&\multirow{2}{*}{NbS$_2$}&\multirow{2}{*}{\cite{Morosin_1974}}\\
$t^+s^+c^-$&3&$R3m$&$3R (\downarrow)$\\ 
$t^+s^-c^0$&2&$P\bar6m2$&$2H_b$&&Nb$_{1+x}$Se$_2$& \cite{Katzke_2004}\\
$t^-s^+c^0$&6&$R\bar3m$&$6R^{\prime\prime}$\\
$t^-s^-c^+$&2&$P6_3/mmc$&$2H_a$&&NbSe$_2$&\cite{Brown_1965}\\
$t^-s^-c^-$&2&$P6_3/mmc$&$2H_c$&&MoS$_2$&\cite{Dickinson_1923}\\ \hline
\end{tabular}
\end{table}


\section{{{\it MX}$_2$} Phase space}

\begin{figure}[b]
\begin{center}
\includegraphics{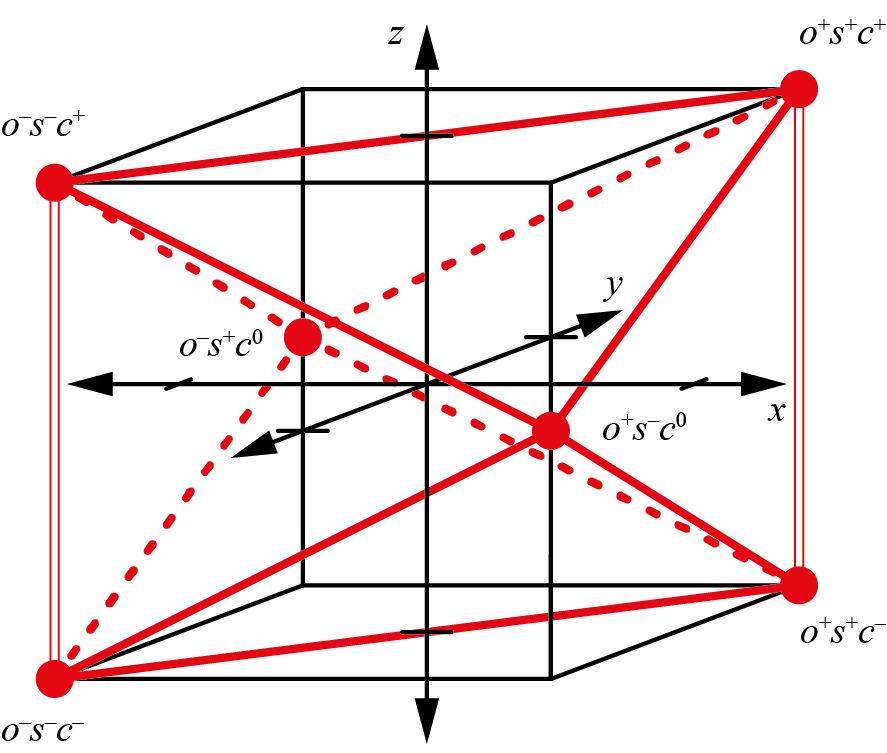}
\end{center}
\caption{\label{fig2} Representation of the phase space accessible to octahedrally-coordinated {\it MX}$_2$ polytypes. Axes are as defined in Eqs.~\eqref{coord1}--\eqref{coord3}; tick-marks denote the correlation function limits of $\pm1$. The six limiting octahedral polytypes listed in Table~\ref{table1} correspond to points on the boundary of the diagram, shown here as red circles.  Every possible octahedral polytype maps to a single point contained within the octahedral volume bounded by these six vertices. The volume is dense in that, for each point within the allowed octahedral volume, there exists at least one polytype (and possibly very many) with the corresponding nearest-neighbour layer and slip correlations. The edges $(1,1,z)$ and $(-1,-1,z)$ (represented here as open lines) are the only exceptions: polytypes exist only for $z=\pm1$ in each case. The phase space accessible to trigonal-prismatic {\it MX}$_2$ polytypes is equivalent in all respects, but with the `$o$' component of the Glazer labels replaced by `$t$'.}
\end{figure}

This labelling approach suggest a means by which the phase space accessible to {\it MX}$_2$ polytypes might be visualised and further understood. In particular, we can create a 3D Cartesian phase space for each of the coordination types (octahedral and trigonal prismatic), where the $x$, $y$, and $z$ components correspond to average nearest-neighbour layer, slip, and coupling correlations, respectively. By letting $l_i$ and $s_i$ denote, for an arbitrary polytype, the handedness of the $i^{\rm th}$ {\it MX}$_2$ layer and of the slip direction in the $i^{\rm th}$ van der Waals gap, taking values of $\pm1$ accordingly, we obtain the Cartesian coordinates
\begin{eqnarray}
x&=&\langle l_il_{i+1}\rangle,\label{coord1}\\
y&=&\langle s_is_{i+1}\rangle,\\
z&=&\langle l_is_i\rangle\label{coord3},
\end{eqnarray}
where the averages are taken over all layers $i$. The relationship to our Glazer notation is most easily seen by considering limiting values of these coordinates. For example, a value $x=1$ corresponds to polytypes in which successive layer orientations are always in-phase, and a value $y=-1$ to those in which successive slip directions are always out-of-phase. For all polytypes---irrespective of their complexity or whether a corresponding Glazer label exists---the values of these coordinates are bounded $-1\leq x,y,z\leq+1$. Enumerating all $4^N$ possible polytypes for finite $N$, it is possible to show that the corresponding coordinates lie within an octahedral volume of Cartesian space [Fig.~\ref{fig2}].  The coordinates lie on a discrete mesh, but this mesh becomes increasingly dense as $N$ increases to form a continuous volume in the limit $N\rightarrow\infty$. The six vertices of this volume correspond to the six Glazer polytypes enumerated in Table~\ref{table1} (for either octahedral or trigonal-prismatic coordination, as appropriate). The mapping is straightforward (by construction) since one simply uses the Glazer indices to extract the corresponding phase-space coordinates. For example, the polytype $o^-s^+c^0$ maps to the coordinate $(-1,+1,0)$.

The coordinates defined in Eqs.~\eqref{coord1}--\eqref{coord3} are simply the nearest-neighbour terms from the correlation functions generated by 1D Ising models of polytype sequences. In this context, they are closely related to the Warren--Cowley parameters used to define correlations in many types of disordered materials \cite{Cowley_1950}. One expects a strong sensitivity to these parameters in the diffraction signature of polytypes, since they are related to the leading Fourier component of any expansion of the scattering function in terms of layer orientations and positions \cite{Welberry_2022}. Similar representations of stacking-fault phase fields have been developed for water ice \cite{Malkin_2015,Playford_2018} and diamond \cite{Salzmann_2015}; in both cases, the representations were two-dimensional, and termed ``stackograms''.

\begin{figure*}
\begin{center}
\includegraphics{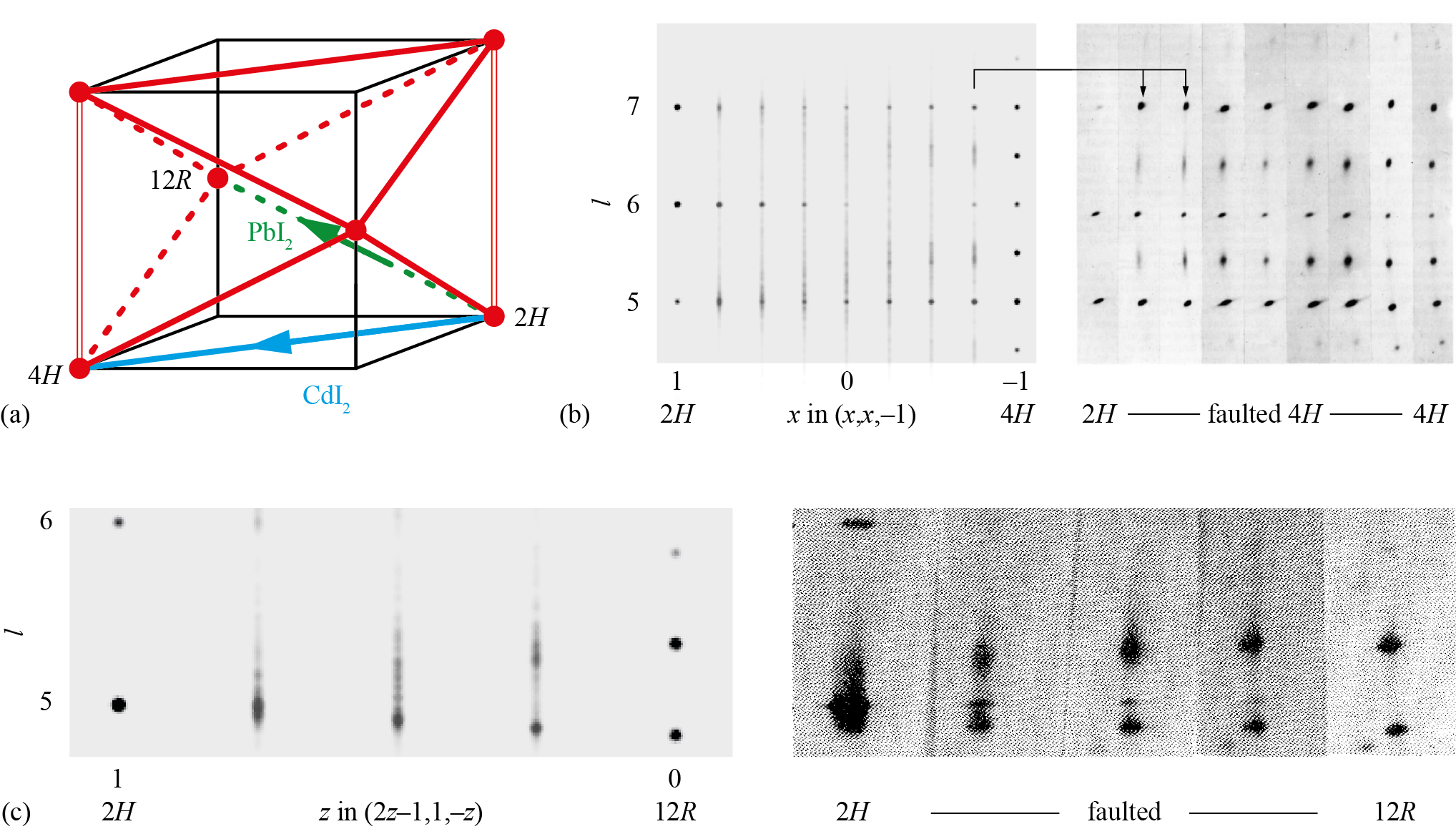}
\end{center}
\caption{\label{fig3} Thermal transitions in {\it MX}$_2$ solids as trajectories in phase space. (a) Phase-space trajectory for the $2H$--$4H$ transition in CdI$_2$ (blue) and the $2H$--$12R$ transition in PbI$_2$ (green). (b) The left-hand panels show $[10l]^\ast$ sections of the single-crystal X-ray diffraction patterns for CdI$_2$ polytypes generated at regular intervals along the relevant trajectory shown in (a). Each strip represents an average over patterns generated for five distinct realisations, each containing 48 layers. Note the continuous evolution of scattering intensity between the $2H$ and $4H$ polytype extrema. The right-hand panels show the experimental data of \cite{Minagawa_1978} for the $2H$ polytype (left), the $4H$ polytype (right) and faulted $4H$ polytypes of varying degrees of disorder (in between). There is good qualitative similarity between some of these patterns and that calculated for the coordinate $(-0.75,-0.75,-1)$. (c) The left-hand panel shows an equivalent progression in $[10l]^\ast$ scattering intensity for PbI$_2$ polytypes generated at regular intervals along the corresponding trajectory shown in (a). The right-hand panel shows the experimental measurements of \cite{Minagawa_1981} for gel-grown PbI$_2$ crystals subjected to successively prolonged heat treatments.}
\end{figure*}

A key advantage of the coordinate representation is that the values of $x,y,z$ are well defined and bounded for all stacking sequences---whether ordered or disordered, thermodynamic or metastable: every possible sequence maps to a single point within the bounded phase volume shown in Fig.~\ref{fig2}. This mapping is not, in general, uniquely reversible. Different stacking sequences can---and often will---map onto the same coordinates.  Stacking sequences with the same coordinates share the same first-order components in the Fourier expansion of their scattering functions (or lattice energy) in terms of layer orientations/positions, and differ only in higher-order components. One important example concerns the origin $x=y=z=0$, which represents an infinity of phases for which the nearest-neighbour pair correlations vanish: these including random stacking, many ANNNI ground-states, and periodic stacking sequences whose ordering wave-vectors lie within the Brillouin zone interior. By contrast, the vertices of the {\it MX}$_2$ phase space are uniquely invertible: for each vertex, the only corresponding polytype is that represented by the relevant Glazer symbol.

There is one oddity of the {\it MX}$_2$ phase space which deserves brief comment: namely that the volume is continuous at all $x,y,z$ except at the edges $(-1,-1,z)$ and $(1,1,z)$, which are forbidden for $z\ne\pm1$. These discontinuities are straightforwardly rationalised in terms of the definitions in Eqs.~\eqref{coord1}--\eqref{coord3}.

Glazer notation is useful in demonstrating the relationships between different structure types. In the context of {\it MX}$_2$ polytypes it is clear, for example, that the $t^+s^-c^0$ and $t^+s^+c^+$ structures comprise trigonal prismatic layers in the same relative orientation, but which slip with different periodicities. We would argue that this point is not at all clear from the conventional labels $2H_b$ and $3R$. The relationships between polytypes become evident geometrically in our Cartesian representation of polytype phase space. Traversing the $x$ direction, for example, corresponds to varying layer-orientation correlations; traversing $y$ corresponds to varying slip-direction correlations. Consequently, we anticipate that transitions from one polytype structure to another---\emph{e.g.}\ as a function of composition, intercalation, temperature, or pressure---may often correspond to continuous trajectories within the  volume of {\it MX}$_2$ phase space.

\section{(Re)interpretation of experiment}

\begin{figure}
\begin{center}
\includegraphics{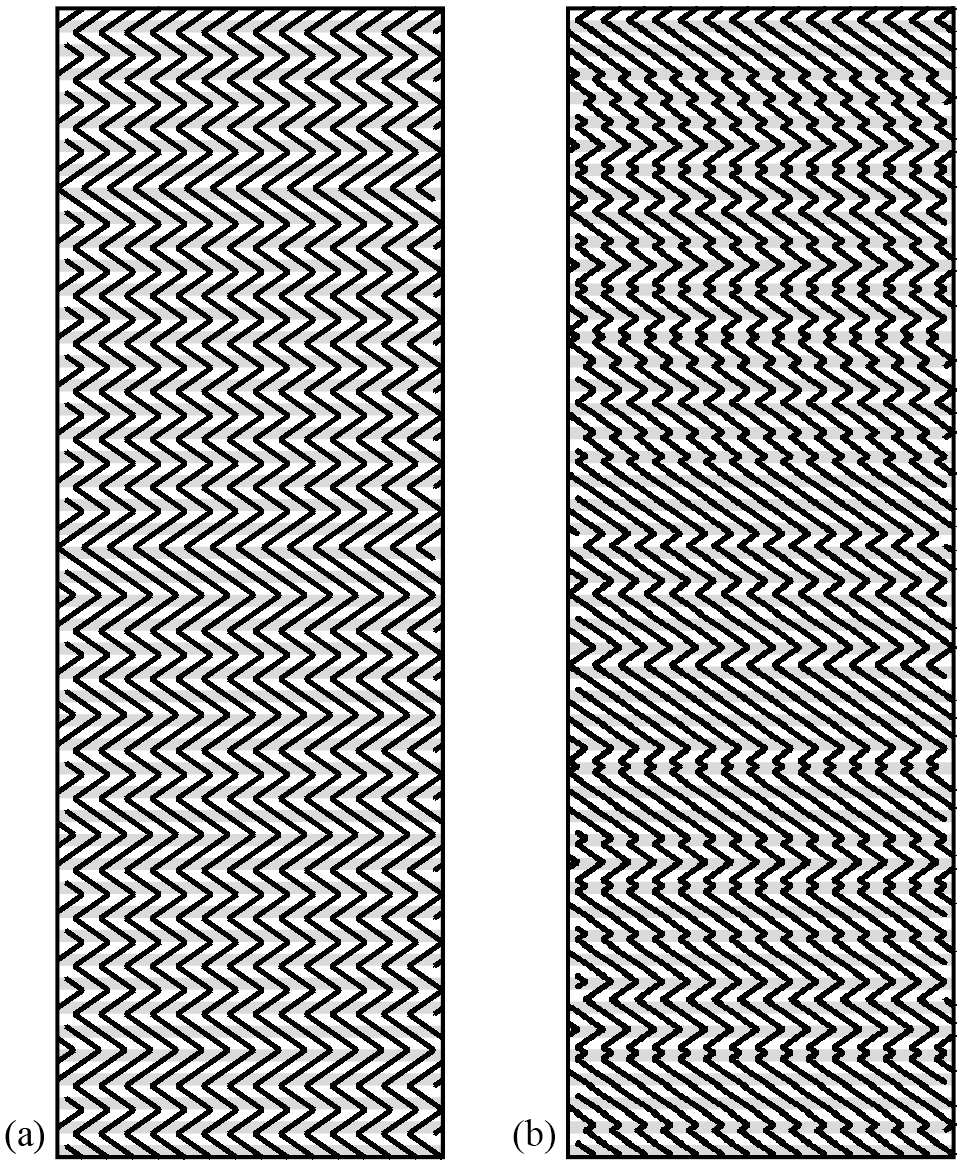}
\end{center}
\caption{\label{fig4} Representative polytype stacking sequences for (a) faulted $4H$-CdI$_2$ and (b) Re$_{0.23}$Ta$_{0.77}$Se$_2$.}
\end{figure}

As a simple example of interpreting the {\it MX}$_2$ phase space, we consider the $2H$ (or $1T$)--$4H$ transition of CdI$_2$ observed on heating \cite{Minagawa_1978}. This transition involves a change from a polytype with Glazer symbol $o^+s^+c^-$ to one with symbol $o^-s^-c^-$, and might be understood as traversing an edge of the phase-space octahedron [Fig.~\ref{fig3}(a)]. Consequently, we generated structural models of polytypes whose stacking sequences correspond to appropriate intermediate coordinates $(x,x,-1)$ ($1>x>-1$), using CdI$_2$ geometries and interlayer spacings commensurate with the ordered $2H$ and $4H$ polytypes. The simulated single-crystal X-ray scattering functions along the $[10l]^\ast$ axis are shown in Fig.~\ref{fig3}(b), where they are also compared against the experimental data of Ref.~\citenum{Minagawa_1978}. We observe qualitatively similar trends between the two. The diffuse streaks observed at intermediate $x$ were interpreted in Refs.~\citenum{Minagawa_1978,Minagawa_1977} as arising from stacking faults, which is entirely consistent with our own analysis: we illustrate in Fig.~\ref{fig4}(a) a representative stacking arrangement that emerges from our $x=-0.75$ configuration.

Disordered stacking arrangements were also reported for PbI$_2$ in a study of its thermally-driven $2H$--$12R$ transition. This transition again corresponds to traversing an edge of the {\it MX}$_2$ phase-space octahedron, now with intermediate coordinates of the form $(2z-1,1,-z)$ $(1>z>0)$. Following a similar process to that outlined above for CdI$_2$, we generated diffraction patterns for intermediate polytypes. In Fig.~\ref{fig3}(c) we compare slices of these diffraction patterns with those reported in Ref.~\citenum{Minagawa_1981}, observing particularly close correspondence between the two.

\begin{figure}[b]
\begin{center}
\includegraphics{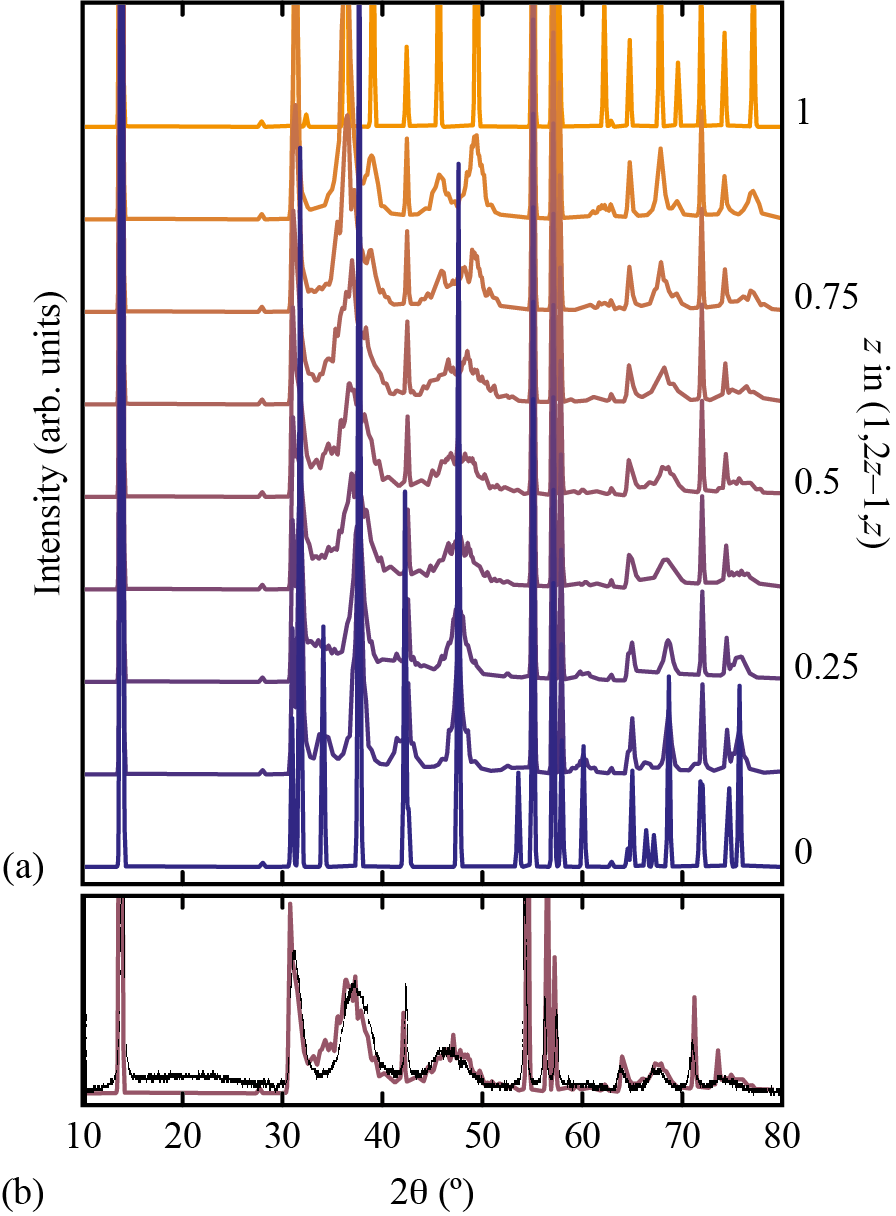}
\end{center}
\caption{\label{fig5} Powder X-ray diffraction (PXRD) patterns relevant to the Re$_x$Ta$_{1-x}$Se$_2$ solid-solution. ReSe$_2$ adopts the CdCl$_2$ structure with Glazer symbol $o^+s^+c^+$, and TaSe$_2$ the $2H_b$ structure with symbol $t^+s^-c^0$. (a) PXRD patterns ($\lambda=1.541$\,\AA) calculated for polytypes generated between these two extremes---\emph{i.e.}\ lying along the $(1,2z-1,z), 0<z<1$ line in the {\it MX}$_2$ phase space. Each pattern is generated from 10 independent stacking sequences, themselves containing 48 layers, of which fractions of $z$ and $1-z$ contained octahedral and trigonal-prismatic coordination, respectively. (b) A comparison of the $z=0.5$ PXRD trace (burgundy line) with the experimental pattern reported in Ref.~\citenum{Hayashi_1995} (black line) for Re$_{0.23}$Ta$_{0.77}$Se$_2$.}
\end{figure}

As a third example, we consider the compositional solid-solution between TaSe$_2$ and ReSe$_2$. The former crystallises in the $3R$ structure type with Glazer symbol $o^+s^+c^+$ \cite{Brown_1965}, and the latter in the $2H_b$ structure type $t^+s^-c^0$ \cite{Alcock_1965}. So, in terms of polytypism, the structural effect of doping Ta for Re is clearly extreme, and involves switching coordination geometry, and also inverting the periodicity of slip direction from in-phase to out-out-phase. One might then expect intermediate compositions to contain a mixture of layer-geometry, slip directions and layer--slip correlations, and hence to correspond to polytypes with Cartesian coordinates along the line $(1,2z-1,z)$ ($1>z>0$). We calculated the diffraction patterns for a series of structural models based on the corresponding polytype correlation functions, and interpolating cell dimensions and internal coordinates from the endmember structures. These X-ray powder diffraction patterns are shown in Fig.~\ref{fig5}(a); what is immediately clear is the evolution of one structure type to another via phases with substantial structured diffuse scattering. Such scattering is evident in experimental data of intermediate-composition phases, and we include in Fig.~\ref{fig5}(b) the experimental powder X-ray diffraction pattern of Re$_{0.23}$Ta$_{0.77}$Se$_2$ \cite{Hayashi_1995} for comparison. We find that this pattern is reasonably well accounted for by a polytype located at $(1,0,0.5)$; a corresponding representative stacking sequence is shown in Fig.~\ref{fig4}(b).

So it seems that, at least for these three examples, physical transformations in {\it MX}$_2$ solids---through variation in temperature or composition---might reasonably be understood in terms of trajectories through the phase space developed in our study. Intermediate states (unsurprisingly) correspond to faulted structures whose diffraction patterns contain textured streaks of diffuse scattering. Revisiting the early crystallographic characterisation of polytypes (\emph{e.g.}\ \cite{Palosz_1983c}) from this new perspective, we might argue it is probably more meaningful to consider \emph{e.g.}\ the 150+ reported polytypes of CdI$_2$ as examples taken from a continuum of states whose structures correspond to polytypes either within the interior or on the boundaries of the {\it MX}$_2$ phase space. We expect there may be many instances in which similar reevaluation is warranted.

\section{Concluding remarks}

If varying composition, on the one hand, and varying temperature, on the other hand, each lead to exploration of the {\it MX}$_2$ phase space along continuous trajectories, then in principle the entire composition/temperature phase diagrams of {\it MX}$_2$ systems lie buried as two dimensional surfaces within the three-dimensional phase space. In favourable cases, one might hope to recover these surfaces from knowledge of the coordinates corresponding to the endmember polytypes at low- and high-temperatures. We hope to explore this possibility in future studies.

Powder X-ray diffraction remains the most straightforward experimental method for characterising the structures of most {\it MX}$_2$ phases. Now that there exist efficient methodologies for calculating diffraction patterns for large stacking models, and subsequently refining the internal parameters from which they are derived (\emph{e.g.}\ layer structure and composition) \cite{Coelho_2016}, it should be possible to locate the effective Cartesian coordinates describing the {\it MX}$_2$ stacking sequence in a sample uniquely from a suitable measurement of its X-ray and/or neutron powder diffraction pattern \cite{Ehrling_2021}.

The approach we have taken here might be straightforwardly extended to other families of layered materials. Layered double hydroxides (LDHs) are closely related to the {\it MX}$_2$ systems on which we have focussed, and these systems support a similarly rich polytypism in practice \cite{Thomas_2006,Leonteva_2020}. Many of the applications of LDHs involve intercalation of guest species within the interlayer regions, and this process can involve changes in polytype. The same is true of {\it MX}$_2$ materials, of course, and one structurally important consequence of intercalation is that it can drive coalignment of the anion positions on either side of the layer gap---a feature intentionally excluded in our model. The {\it AMX}$_2$ delafossite structures are an obvious limiting example, and the Glazer notation would need to be adapted to account for these additional possibilities. Likewise, layered polytypes based on \emph{e.g.}\ honeycomb BI$_3$ layers or square-grid FeSe layers may be interpretable using conceptually similar approaches---but there will be meaningful differences. A recurring distinction between polytypism in layered materials and that in dense phases (\emph{e.g.}\ SiC) is that the former always requires independent consideration of correlations between successive layers, correlations between successive van der Waals gaps, \emph{and} correlations between layers and gaps. From a statistical mechanical viewpoint, the problem is one of coupled 1D chains, which is presumably why the behaviour of layered materials is not necessarily well captured by conventional (single-chain) 1D models.

The ultimate goal of establishing composition--polytype relationships in various layered {\it MX}$_2$ systems is to establish methodologies for synthetic control over polytype selection. Doing so will provide a means for targeting polytypes with particular features of interest---\emph{e.g.}\ preservation or breaking of inversion symmetry, or chirality, or polarity. Our hope in presenting a spatial representation of the {\it MX}$_2$ phase field is to highlight relationships between different polytypes that might not have been obvious from established nomenclatures. Such relationships motivate the use of synthesis parameters (composition, temperature, pressure) to navigate {\it MX}$_2$ phase space in a controlled manner. Establishing clear polytype--property relationships then provides a means for linking structural complexity to material function in {\it MX}$_2$ systems.

\section*{Appendix}
We provide crystallographic details of the twelve ordered polytypes listed in Table~\ref{table1}. In each case, the crystal symmetry and atom coordinates were determined using the {\sc findsym} code \cite{Stokes_2005}. The symbols $a_{\rm ref}$ and $c_{\rm ref}$ denote, respectively, the in-plane hexagonal cell constant and the stacking distance. The parameter $z$ denotes the fractional out-of-plane coordinate for the {\it X} anions relative to the {\it M} layer.

Glazer symbol $o^+s^+c^+$: $R\bar3m$; $a=a_{\rm ref}, c=3c_{\rm ref}$; {\it M} $3a (0,0,0)$; {\it X} $6c (0,0,\frac{1}{3}-z)$.

Glazer symbol $o^+s^+c^-$: $P\bar3m1$; $a=a_{\rm ref}, c=c_{\rm ref}$; {\it M} $1a (0,0,0)$; {\it X} $2d (\frac{1}{3},\frac{2}{3},-z)$.

Glazer symbol $o^+s^-c^0$: $R\bar3m$; $a=a_{\rm ref}, c=6c_{\rm ref}$; {\it M} $6c (0,0,\frac{1}{12})$; {\it X}1 $6c (0,0,\frac{5}{12}-\frac{z}{6})$; {\it X}2 $6c (0,0,\frac{3}{4}+\frac{z}{6})$.

Glazer symbol $o^-s^+c^0$: $R\bar3m$; $a=a_{\rm ref}, c=6c_{\rm ref}$; {\it M}1 $3a (0,0,0)$; {\it M}2 $3b (0,0,\frac{1}{2})$; {\it X}1 $6c (0,0,\frac{1}{3}-\frac{z}{6})$; {\it X}2 $6c (0,0,\frac{1}{6}-\frac{z}{6})$.

Glazer symbol $o^-s^-c^+$: $P6_3mc$; $a=a_{\rm ref}, c=2c_{\rm ref}$; {\it M} $2b (\frac{1}{3},\frac{2}{3},\frac{1}{2})$; {\it X}1 $2a (0,0,\frac{z}{2})$; {\it X}2 $2b (\frac{1}{3},\frac{2}{3},-\frac{z}{2})$.

Glazer symbol $o^-s^-c^-$: $P6_3mc$; $a=a_{\rm ref}, c=2c_{\rm ref}$; {\it M} $2b (\frac{1}{3},\frac{2}{3},\frac{1}{2})$; {\it X}1 $2b (\frac{1}{3},\frac{2}{3},\frac{z}{2})$; {\it X}2 $2a (0,0,-\frac{z}{2})$.

Glazer symbol $t^+s^+c^+$: $R3m$; $a=a_{\rm ref}, c=3c_{\rm ref}$; {\it M} $3a (0,0,0)$; {\it X}1 $3a (0,0,\frac{z}{3}-\frac{1}{3})$; {\it X}2 $3a (0,0,\frac{2}{3}-\frac{z}{3})$.

Glazer symbol $t^+s^+c^-$: $R3m$; $a=a_{\rm ref}, c=3c_{\rm ref}$; {\it M} $3a (0,0,0)$; {\it X}1 $3a (0,0,\frac{1}{3}-\frac{z}{3})$; {\it X}2 $3a (0,0,\frac{z}{3}-\frac{2}{3})$.

Glazer symbol $t^+s^-c^0$: $P\bar6m2$; $a=a_{\rm ref}, c=2c_{\rm ref}$; {\it M}1 $1e (\frac{2}{3},\frac{1}{3},0)$; {\it M}2 $1b (0,0,\frac{1}{2})$; {\it X}1 $2g (0,0,-\frac{z}{2})$; {\it X}2 $2h (\frac{1}{3},\frac{2}{3},\frac{1}{2}-\frac{z}{2})$.

Glazer symbol $t^-s^+c^0$: $R\bar3m$; $a=a_{\rm ref}, c=6c_{\rm ref}$; {\it M} $6c (0,0,-\frac{1}{12})$; {\it X}1 $6c (0,0,\frac{1}{4}-\frac{z}{6})$; {\it X}2 $3a (0,0,\frac{1}{4}+\frac{z}{6})$.

Glazer symbol $t^-s^-c^+$: $P6_3/mmc$; $a=a_{\rm ref}, c=2c_{\rm ref}$; {\it M} $2b (0,0,\frac{1}{4})$; {\it X} $4f (\frac{1}{3},\frac{2}{3},\frac{1}{4}+\frac{z}{2})$.

Glazer symbol $t^-s^-c^-$: $P6_3/mmc$; $a=a_{\rm ref}, c=2c_{\rm ref}$; {\it M} $2c (\frac{1}{3},\frac{2}{3},\frac{1}{4})$; {\it X} $4f (\frac{1}{3},\frac{2}{3},\frac{3}{4}+\frac{z}{2})$.

\begin{acknowledgments}
The authors gratefully acknowledge financial support from Hertford College, Oxford to E.H.W. and from the European Research Council to A.L.G. (Grants 279705, 788144). A.L.G. thanks Shintaro Ishiwata (Osaka) for useful discussions.
\end{acknowledgments}

\end{document}